\documentclass[%
 aip,
 amsmath,amssymb,
 reprint,%
]{revtex4-1}

\usepackage{graphicx}
\usepackage{dcolumn}
\usepackage{bm}

\usepackage[utf8]{inputenc}
\usepackage[T1]{fontenc}
\usepackage{mathptmx}
\usepackage{etoolbox}

\usepackage{lineno}

\makeatletter
\def\@email#1#2{%
 \endgroup
 \patchcmd{\titleblock@produce}
  {\frontmatter@RRAPformat}
  {\frontmatter@RRAPformat{\produce@RRAP{*#1\href{mailto:#2}{#2}}}\frontmatter@RRAPformat}
  {}{}
}%
\makeatother

\begin{document}

\preprint{AIP/123-QED}

\title{Rapid muon tomography for border security}
\author{Anzori Sh.~Georgadze}
\email{a.sh.georgadze@gmail.com}
\affiliation{%
Institute for Nuclear Research, National Academy of Sciences of Ukraine, Prospekt Nauky 47, 03680 Kyiv, Ukraine
}%

\begin{abstract}
Cosmic-ray muon tomography is a promising technique for border security applications, leveraging highly penetrating cosmic-ray muons and their interactions with various materials to generate 3D images of large and dense objects, such as shipping containers. Using scattering and absorption of muons as they pass through dense cargo materials, muon tomography provides a viable solution for customs and border security by enabling the verification of shipping container declarations and preventing illegal trafficking.
In this study, we utilized Monte Carlo simulations to evaluate the effectiveness of muon tomography for cargo characterization and contraband detection in various smuggling scenarios. Our results demonstrate that muon tomography can offers a novel approach to cargo inspection, moving beyond traditional 3D image reconstruction. Instead, it analyzes muon scattering and absorption rates in real time during scanning, enabling the prompt detection of discrepancies between actual cargo contents and declared goods within just 10 to 20 seconds.
This method is particularly effective for cargo consisting of uniform loads composed of a single material or product, a common practice in shipping. Unlike traditional X-ray radiography, which analyzes detailed 2D images, muon tomography begins evaluating scatter-absorption rates within the first few seconds of scanning. This early assessment enables cargo evaluation long before a statistically reliable 3D image is formed, significantly improving scanning throughput without disrupting trade flow.
\end{abstract}

\maketitle

\section{Introduction}
Muon tomography is an emerging technique with promising applications in various fields such as non-destructive testing, underground cavities, archaeology, glaciers ~\cite{tanaka2007high, lesparre2012density, morishima2017discovery, saracino2017imaging, nishiyama2017first}. 

As global trade continues to expand, the challenge of intercepting illicit smuggling activities at borders has become increasingly complex. Traditional cargo inspection methods often struggle to keep pace with sophisticated concealment techniques employed by smugglers. 
Muon tomography has been proposed as a tool for border security applications~\cite{schultz2003cosmic, bonechi, barnes2023cosmic, borozdin2003, explosives, yifan2018discrimination, aastrom2016precision, checchia2016review, antonuccio2017muon, pugliatti2014design, morris2013new, preziosi2020tecnomuse, lowZ, georgadze2023geant4, georgadze2024simulation, georgadze2024, preziosi2020tecnomuse, georgadze2023geant4, rapidcargo, Anzori_Auto}. 
This technology, recently proposed to cargo inspection, offers unique capabilities. It provides exceptional penetration for effectively screening even the densest materials and cargo, surpassing the capabilities of traditional X-ray systems. Additionally, muon tomography is non-hazardous to humans and requires no shielding, allowing scanning sealed cargo without opening containers. The European project, "Cosmic Ray Tomograph for Identification of Hazardous and Illegal Goods hidden in Trucks and Sea Containers" (SilentBorder)~\cite{sbwebsite}, focuses on the development and in-situ testing of a high-technology scanner designed for border guards, customs, and law enforcement authorities to inspect shipping containers at border control points.

In this paper, several Monte Carlo (MC) simulation studies were conducted to evaluate the muon tomography method and the algorithms used for image reconstruction in container transshipment security applications. Various configurations and smuggling scenarios were considered and simulated. The goal was to develop a quantitative method for the rapid assessment of the presence of illegal and hazardous materials concealed within legitimate goods, as well as to create noise reduction techniques for visualizing hidden objects. The statistical method is used to compare the combined analysis of muon scattering and absorption rates for different material configurations obtained from the scan data, enabling the creation of a map for the rapid assessment of customs declarations.
\vspace{-5mm}

\section{Materials and Methods}
We use GEANT4~\cite{AGOSTINELLI} to model various muon tomography scenarios for border security applications. This enables realistic simulations of muon interactions with cargo materials and hidden items. 
When muons interact with matter, they scatter, and by measuring the angles and intensities of these scattered muons it is possible to create 3D images of the interior of objects or structures. The fraction of cosmic muons can be absorbed by the crossed materials. 
Muon absorption depends on the density and composition of the material. 

Cosmic-ray muons are naturally occurring charged particles generated in extensive atmospheric showers when high-energy cosmic rays from space interact with the Earth's atmosphere. Their flux at sea level is approximately 1 cm$^{-2}$ min$^{-1}$, with an average energy of about 3 GeV.

\subsection{The Muon Scattering Tomography Method} 
The scattering muon tomography technique calculates the deflection of muons from their straight trajectory because of multiple coulomb interactions, which in turn depend on cargo density and chemical composition. 
The angular distribution of scattered muon of momentum p is approximately Gaussian, with zero mean and standard deviation given by:
\begin{equation}\label{eq:1}
	\sigma_\theta = \frac{13.6 MeV }{\beta cp}\sqrt{\frac{L}{_{X_{0}}}}(1+0.038)ln\frac{L}{_{X_{0}}})
\end{equation}
where $\beta$ is the ratio between velocity of muon \emph{V} to velocity of light \emph{c}, \emph{X\textsubscript{0}} is the radiation length of the material, \emph{L} is the length of the material traversed. \emph{X\textsubscript{0}} is a material property and depends on the density of the material $\rho$, the atomic mass \emph{A} and the atomic number \emph{Z} and can be expressed as\cite{lynch1991approximations}:
\begin{equation}\label{eq:2}
X_{0} = \frac{716.4g/cm^2}{\rho }\frac{A}{Z(Z+1)ln(287/\sqrt{Z})}
\end{equation}

\subsection{The Muon Absorption Method} 
In the absorption tomography approach, the tracks of muons stopped in the cargo can be reconstructed using an algorithm similar to that described in the publication~\cite{vanini2019muography, Blanpied, rengifo2024design}. 
The absorption muon tomography technique calculates the fraction of muons that were stopped in cargo by linking upper and lower tracking detectors. It focuses on muons detected by the upper detectors but absorbed within the imaging volume before reaching the lower detectors. The algorithm tracks muon paths and counts voxel crossings. 
Reconstructed muon tracks determine the path length $d_{ij}$ of each muon through voxel \textit{j}. 
The stopping power $S_{j}$ represents the muon’s energy loss per unit distance in that voxel. The total absorption $N_{abs,i}$ along a muon’s path is the sum of stopping powers across all traversed voxels:

\begin{equation}\label{3}
N_{abs,i} =\sum_{j}d_{ij}S_{j}
\end{equation}

\begin{figure}[b]
\centering
\includegraphics[width=0.23\textwidth]{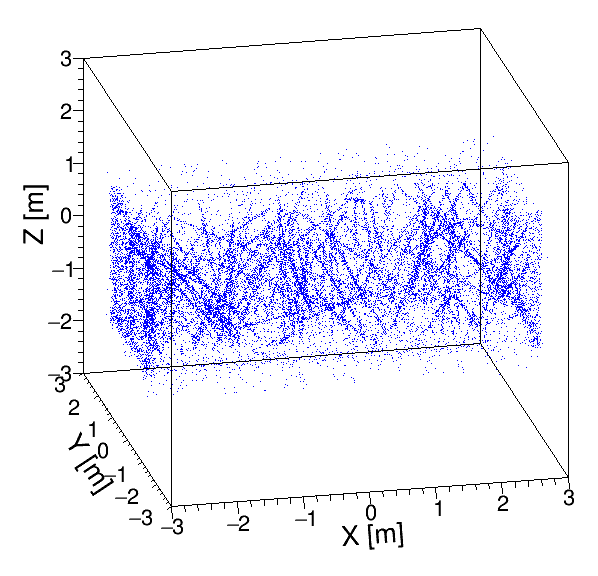}
\caption{ 
Visualizations using GEANT4 of muon tracks (n~=~200) in a shipping container with cargo, generated using the Cosmic Ray Shower Library (CRY)~\cite{hagmann2007cosmic}.}
\label{fig:fig1}
\end{figure}

Muons interacting with the contents of a shipping container can be significantly deflected or even stopped due to their long path lengths, often extending meters, as shown in Figure~\ref{fig:fig1}. 
The extent of muon scattering and absorption depends on the cargo’s density, resulting in distinct deflection and absorption patterns for different materials. This variation provides a viable method for material identification within shipping containers.

\subsection{Cargo Inspection Procedures. } 
The cargo inspection process may follows a two-stage approach. In the first stage, a cargo image is reconstructed within 10–20 seconds of scanning, followed by an combined analysis of the cargo's scattering and absorption rates. These measured rates are then compared with the expected values predicted based on the customs declaration.
If discrepancies are identified, the scanning process continues to create three-dimensional image ensuring statistically significant accuracy. 
Such a two-stage approach enhances the ability to validate inconsistencies and potentially visualize any concealed contraband within the cargo. 
Due to the different sizes and loading configurations of the cargo, in order to correctly verify cargo material object detection techniques have to be first applied to identify the shape and dimensions of the cargo in the reconstructed image of the container~\cite{Anzori_Auto}.

\section{Results}

\subsection{Modeling the Substitution of Declared Goods with Contraband. } 
We consider the substitution of declared clothes with high-value branded jeans. In this scenario, a portion of the declared low-cost cotton jeans is secretly replaced with high-end branded jeans to evade customs duties or import restrictions. This substitution can be detected in muon tomography because of differences in the packaging structure and material density. 

\begin{figure}[b]
\begin{minipage}{1.\linewidth}
\centering
\includegraphics[width=0.8\linewidth]{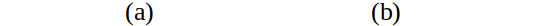}
\vspace{-0.5mm}  
\end{minipage}
\centering
\includegraphics[width=0.2\textwidth]{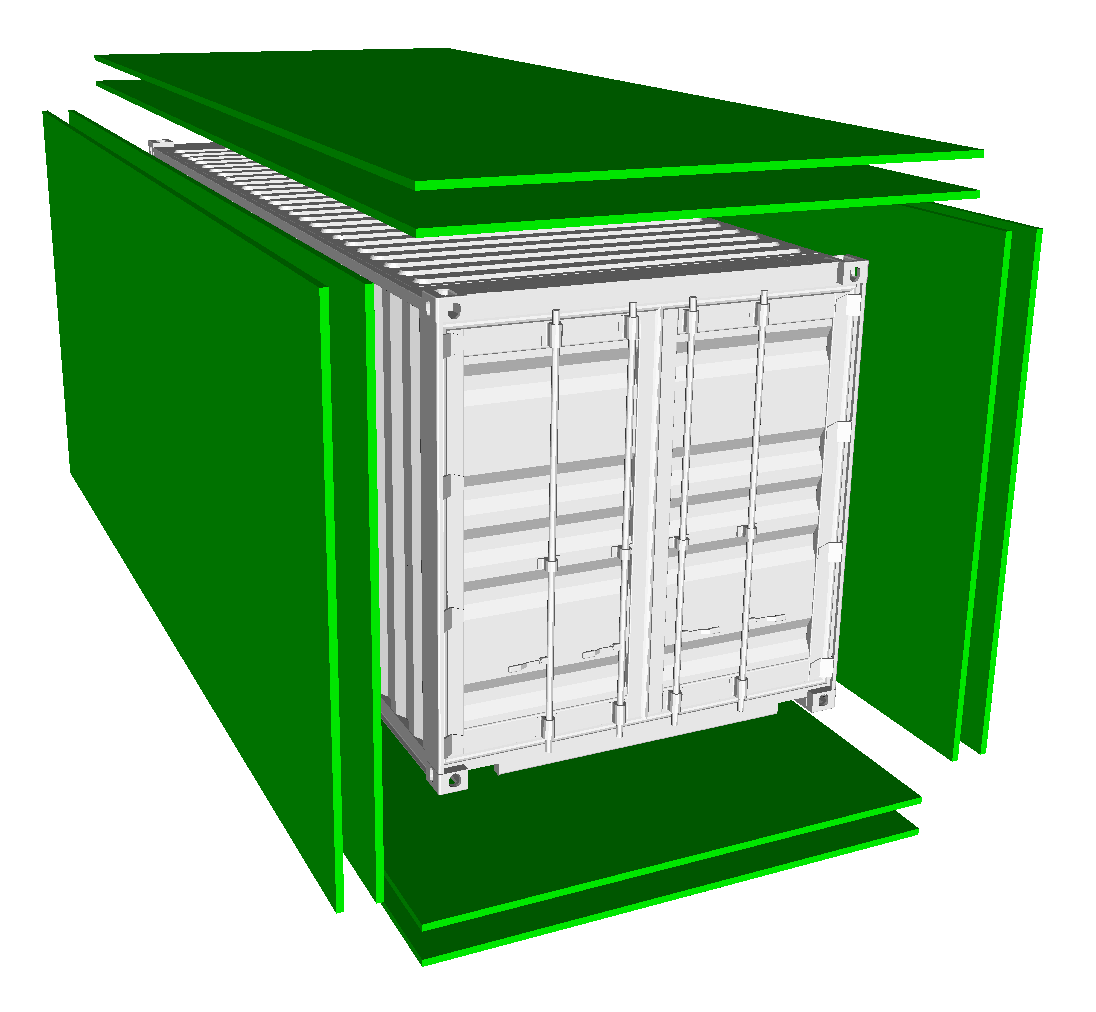}
\includegraphics[width=0.26\textwidth]{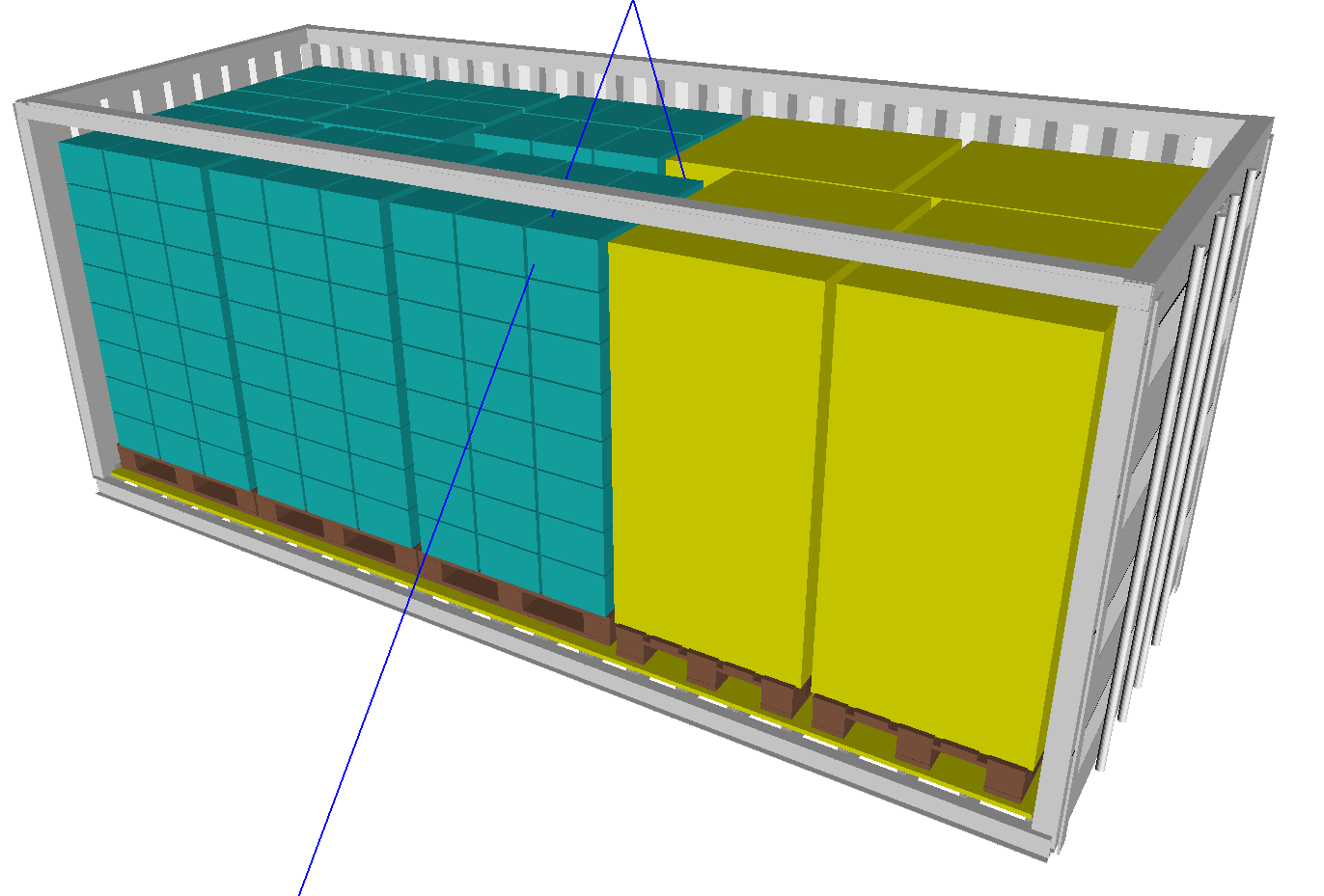}
\caption{(a) Schematic view of cosmic ray tomography
station. 
(b) GEANT4 visualizations of a container with bulk-packed cotton jeans (yellow) and branded jeans (cyan) in individual packaging on pallets.}
\label{fig:fclothes}
\end{figure}

In the GEANT4 simulation, the muon tomography station (MTS) is composed of plane detectors made of plastic scintillator, which have a 100\% detection efficiency. The geometry of the MTS is illustrated in Figure~\ref{fig:fclothes}(a). The MTS consists of tracking modules located at the top, bottom, and sides, with plane detector dimensions of 8 × 4 × 0.001 m$^3$, fully covering the shipping container.
The GEANT4 model of a standard 20-foot shipping container (measuring $ 6.05 \times 2.59 \times 2.43~\text{m}^3$) was filled with cargo on standard pallets, as shown in Figure~\ref{fig:fclothes}(b).
In this figure, bulk-packed cotton jeans are represented in yellow, while branded jeans in individual packaging, placed in boxes on pallets, are shown in cyan.
Cheap cotton jeans were modeled as a textile made of cotton fiber with a bulk density of approximately 0.4 g/cm$^3$, while branded jeans had a bulk density of about 0.2 g/cm$^3$. The difference in bulk density arises from variations in packaging and air content rather than the chemical density of the fabric itself. Cheap jeans are typically packed in bulk, tightly compressed into large cardboard boxes, minimizing air gaps and increasing bulk density. In contrast, branded jeans are often individually wrapped in plastic, placed in sturdier packaging, or folded differently, introducing more air pockets and resulting in a lower bulk density.
Variations in cargo materials influence muon scattering and absorption rates, enabling the detection of discrepancies between uniform and substituted cargo sections. To verify the dependence of scatter-absorption rates on cargo material in this scenario, we simulated 5,000 data samples using the CRY muon generator with 100,000 muons sampled on a 10 m $\times$ 10 m surface. 
This number of muons corresponds to approximately 10 seconds of scan time.
At the top of each 2D histogram, the corresponding 1D histograms for both distributions are displayed. We apply the Point-of-Closest-Approach (PoCA) algorithm \cite{hoch2009muon} to determine the closest point between incoming and outgoing muon tracks and calculate the scattering angle.

\begin{figure}
\begin{minipage}{1.\linewidth}
\centering
\includegraphics[width=0.8\linewidth]{figures/ab.png}
\vspace{-0.5mm}  
\end{minipage}
\centering
\includegraphics[width=0.24\textwidth]{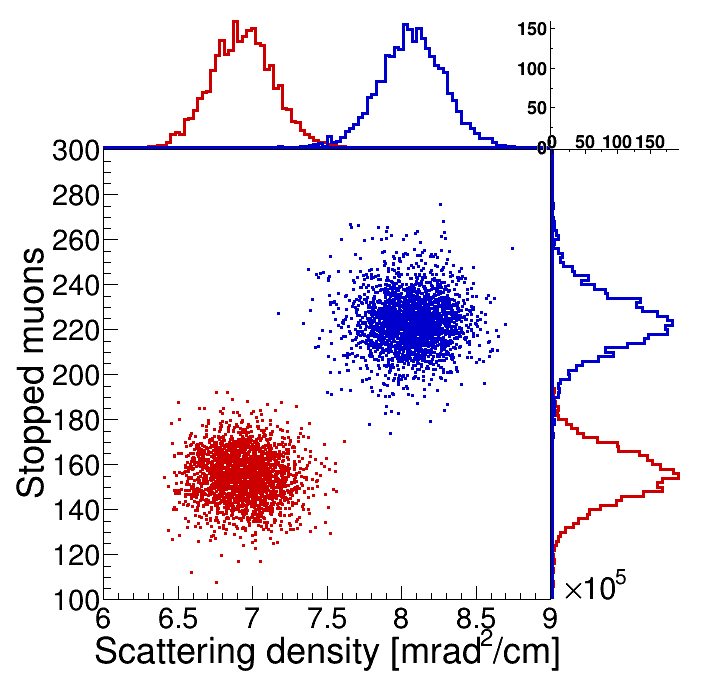}
\includegraphics[width=0.23\textwidth]{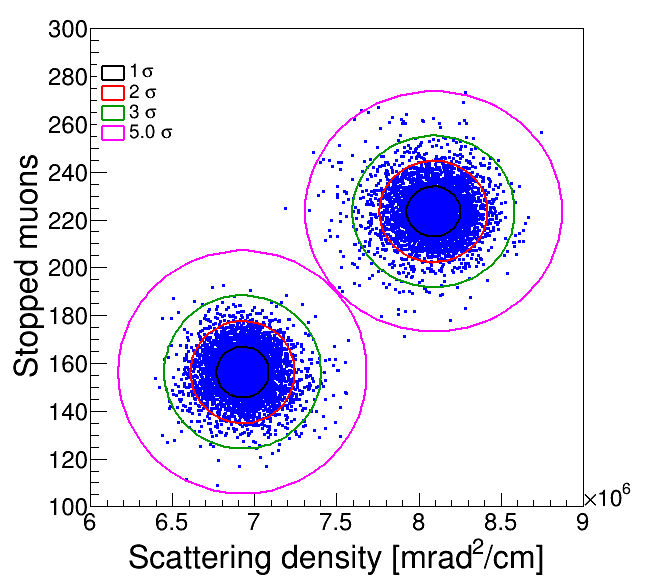}
\caption{(a) The distributions of 5,000 data samples generated for scenario of bulk-packed cotton jeans in container represented in blue, while scenario with partial substitution with branded jeans are shown in red. At the top of each 2D histogram, the corresponding 1D histograms for both distributions are displayed.
(b) The same distributions are fitted with two component 2D Gaussian Mixture Model to quantify the discrimination accuracy between two scenarios. The black, red and green confidence ellipses for each distribution are set to show 1, 2 and 3 $\sigma$ confidence levels (CL). The magenta confidence ellipses show the 5 $\sigma$ CL at which the distributions are discriminated.}
\end{figure}

\begin{figure}[t]
\begin{minipage}{1.\linewidth}
\centering
\includegraphics[width=0.8\linewidth]{figures/ab.png}
\vspace{-0.5mm}  
\end{minipage}
\label{fig:fclothes2}
\includegraphics[width=0.45\textwidth]{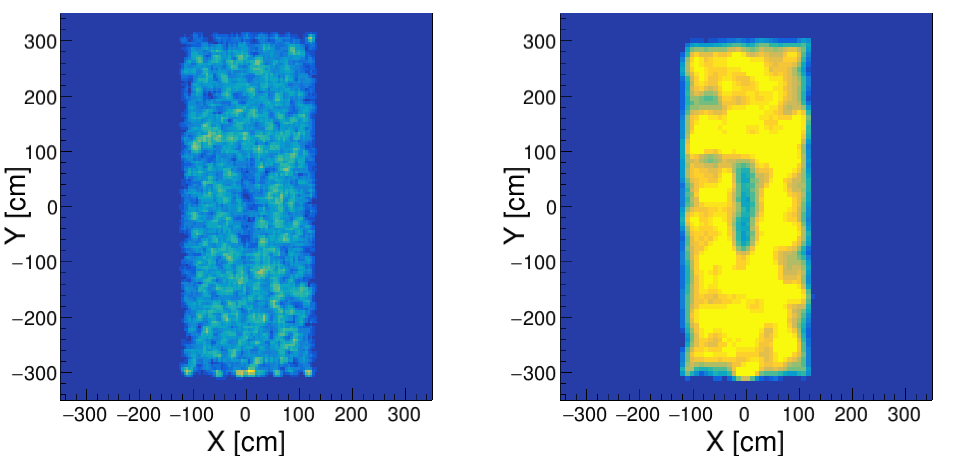}
\begin{minipage}{1.\linewidth}
\centering
\includegraphics[width=0.8\linewidth]{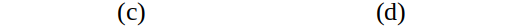}
\vspace{-0.5mm}  
\end{minipage}
\includegraphics[width=0.45\textwidth]{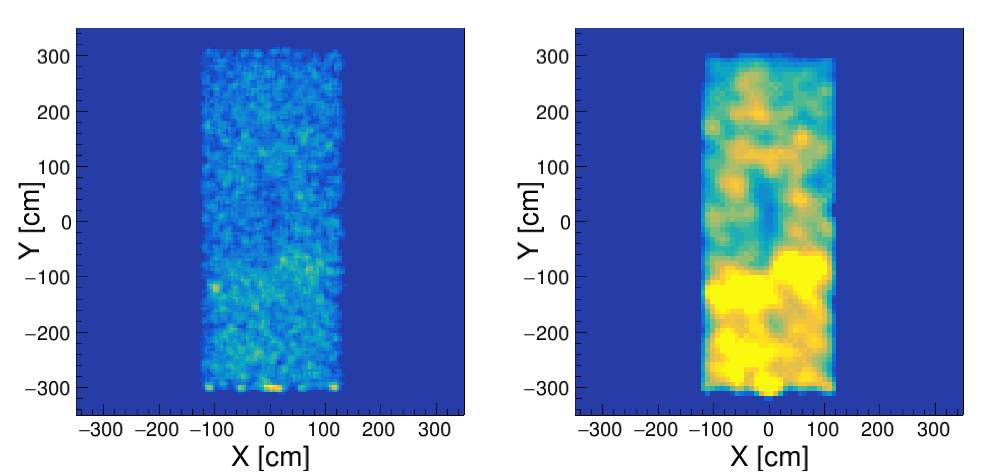}
\caption{Reconstructed XY projections of 3D PoCA images of a container filled with bulk-packaged cotton jeans (a, b) and a container in which some of the pallets are replaced with individually packaged branded jeans (c, d). The simulated images were obtained by generating 300,000 muons using the CRY muon generator, corresponding to a scan time of 30 seconds. In (b) and (d), a threshold was applied during data processing to improve image contrast.}
\label{fig:fclothes3}
\end{figure}
To create images of cargo in this scenario we simulated 300,000 muons sampled on the surface 10 m $\times$ 10 m. This number of muons corresponds to approximately 30 seconds of scan time. 
As shown in the reconstructed images in Figure~\ref{fig:fclothes3}(a and b), the bright yellow color indicates regions of high scattering and absorption of muon tracks, corresponding to bulk-packed cotton jeans (with higher density), while the dark areas are attributed to low-density branded jeans in individual packaging (with lighter density). Processing image created using PoCA method is performed using ROOT package~\cite{ROOT}.
In this scenario, cargo substitution with contraband material can be detected within 10 seconds using statistical analysis of scatter-absorption rates and visualized within 30 seconds.

\subsection{Modeling Hidden Hazardous Materials in Legal Cargo. } 

\begin{figure}[t]
\begin{minipage}{1.\linewidth}
\centering
\includegraphics[width=0.75\linewidth]{figures/ab.png}
\vspace{-0.5mm}  
\end{minipage}
\centering
\includegraphics[width=0.24\textwidth]{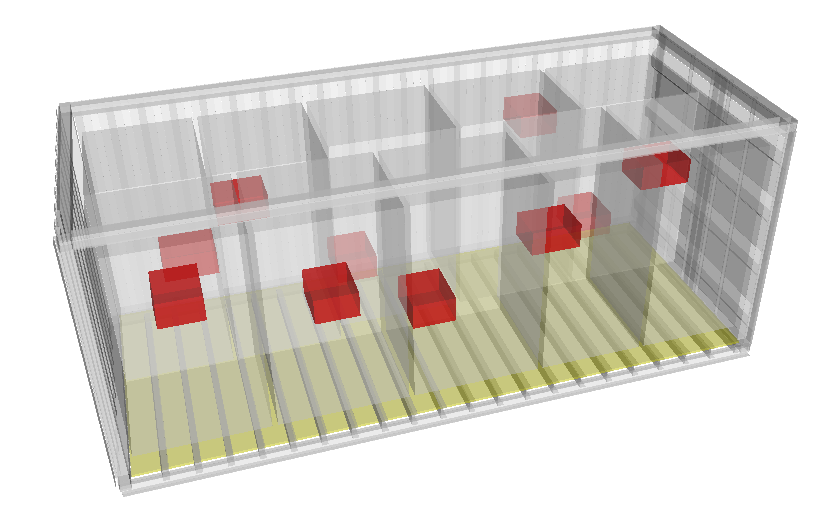}
\includegraphics[width=0.21\textwidth]{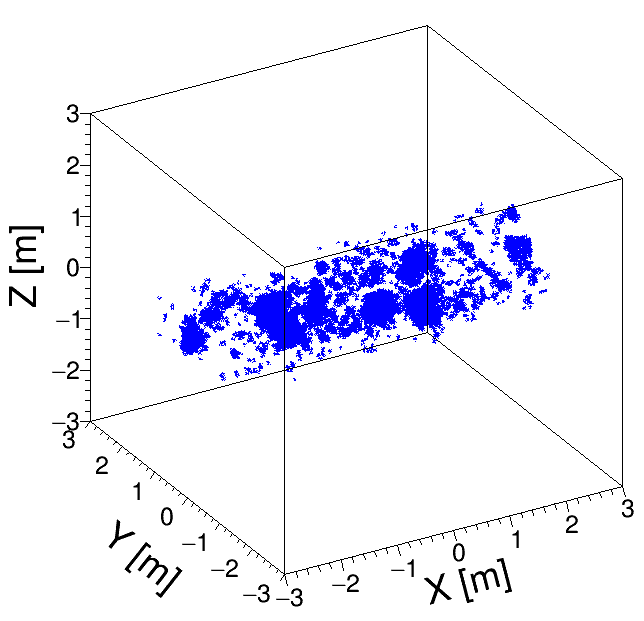}
\caption{(a) 
A GEANT4 simulation of a container where RDX is hidden within dry pasta placed on pallets. (b) Reconstructed 3D image of the container using the PoCA method. Median filtering is applied to the 3D reconstruction to reduce noise and improve visualization of hidden items.}
\label{fig:figure4}
\end{figure}

To simulate a scenario involving a shipping container loaded with cargo containing hidden hazardous materials, we used a GEANT4 model of a standard 20-foot shipping container (Figure~\ref{fig:figure4}a), containing legal cargo arranged in carton boxes placed on pallets. Among these boxes, hazardous materials are concealed. Specifically, one box on each pallet is loaded with explosive material—Royal Demolition eXplosive (RDX), which has a density of 1.812 g/cm$^3$.   
To create a tomographic image of the shipping container, 5 million muons (equivalent to a 5-minute scanning time) were simulated using the CRY muon generator, sampled on a 10 $\times$ 10 m$^2$ surface.
Figure~\ref{fig:figure4}b shows the 3D PoCA reconstruction after applying median filtering to reduce noise.
In order to reject noise from legal cargo determine positions of hidden illegal goods and an estimate of their density besides image filtering we perform spatial cuts, removing PoCA points outside shipping container area. Next we subtract the image of an empty container from the 3D image of a loaded container. 

\begin{figure}[b]
\begin{minipage}{1.\linewidth}
\centering
\includegraphics[width=0.75\linewidth]{figures/ab.png}
\vspace{-0.5mm}  
\end{minipage}
\includegraphics[width=0.25\textwidth]{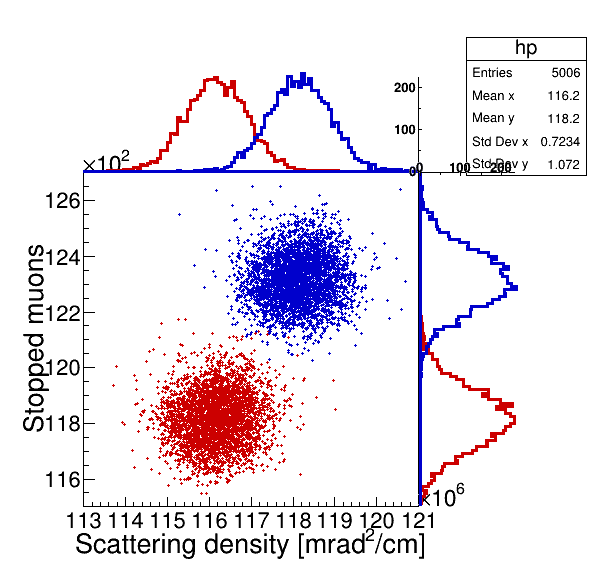}
\includegraphics[width=0.22\textwidth]{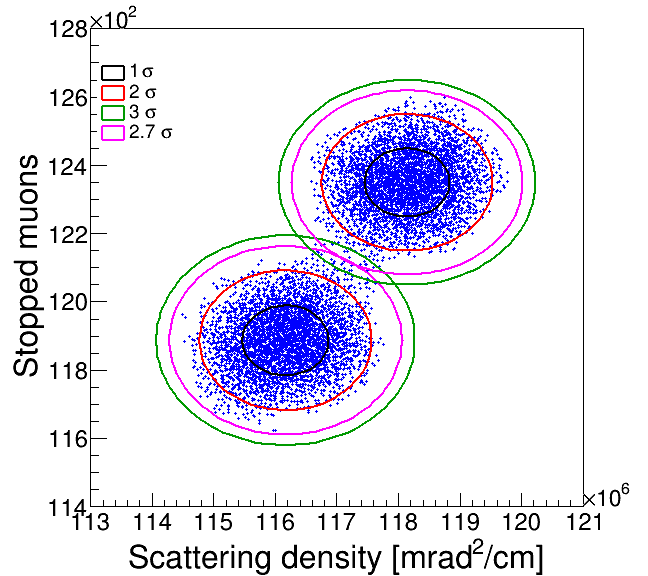}
\caption{(a) Histograms show the distribution of 5000 data samples for legal cargo - dry pasta (red data points) and the distribution of 5000 data samples for legal cargo mixed with 500 kg of RDX explosive randomly distributed in boxes on pallets (blue data points).
(b) Scatter plots of data samples fitted to a two-component 2D Gaussian mixture model. The black, red and green confidence ellipses for each distribution are adjusted to show confidence levels (CL) of 1, 2 and 3 $\sigma$. The magenta confidence ellipses show CL 4.5 $\sigma$, at which the distributions are discriminated.}
\label{fig:f222}
\end{figure}

To verify the applicability of rapid detection methods for the scenario of hidden RDX in cargo, we simulated 100,000 muons, sampled over a 10 m  $\times$  10 m surface area, using the CRY generator. We considered two geometries: one where the container is loaded only with legal cargo (dry pasta, density 0.95 g/cm$^3$), and another where one box on each pallet is replaced with RDX.
The 5 min scanning time images of container with explosive RDX hidden in 5 tones of cargo (clothes) are shown in figure \ref{fig:f222}. 
An advantage of muon tomography - segmentation of 3D image into slices makes possible detection of hidden explosive. 
Figures \ref{fig:f222}(a) and (b) shows results of statistical analysis demonstrating at which CL two scenarios are discriminated.   

\subsection{Simulation of Methamphetamine Concealment within Cement Shipments}

Another important scenario for evaluating the performance of muon tomography involves the concealment of lower-density contraband materials within higher-density cargo. In such cases, the tomographic image reveals the illicit substance as localized regions of reduced scattering density, appearing as voids or anomalies, within the surrounding dense matrix.
A representative real-world example of this scenario is the concealment of methamphetamine hydrochloride powder inside cement bags~\cite{customs}. 
Cement, being a high-density material, can effectively attenuate X-rays and is known to pose significant challenges for conventional radiographic screening techniques. This makes muon tomography particularly advantageous, as it relies on multiple scattering of naturally occurring cosmic muons rather than attenuation, thereby offering greater sensitivity to differences in atomic number ($Z$) and material density.



\begin{figure}[htbp]
\centering
\begin{minipage}{1.\linewidth}
\centering
\includegraphics[width=0.6\linewidth]{figures/ab.png}
\vspace{-1.mm}  
\end{minipage}	
\includegraphics[width=0.22\textwidth]{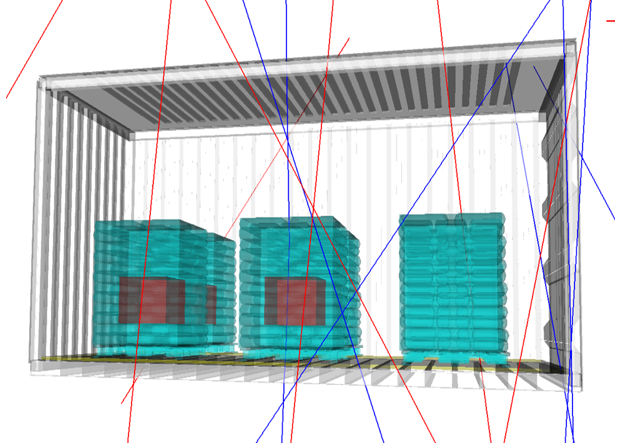} 
\includegraphics[width=0.22\textwidth]{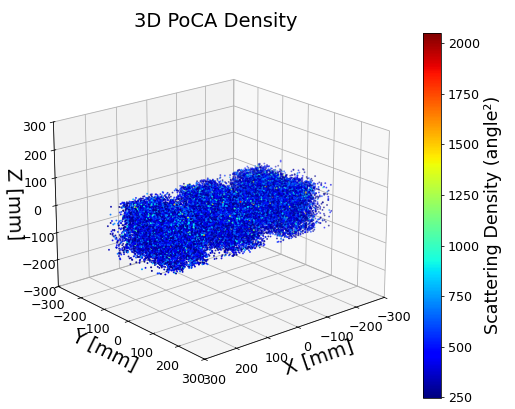} 
\caption{(a) Conceptual illustration of the GEANT4 geometry representing the scenario in which methamphetamine is concealed within cement bags loaded on six pallets inside a 20-foot cargo container; (b) Simulated 3D tomographic reconstruction of concealed methamphetamine within the cement bags.}
\label{fig:cement_meth_geometry}
\begin{minipage}{1.\linewidth}
\centering
\includegraphics[width=0.6\linewidth]{figures/ab.png}
\vspace{-1.mm}  
\end{minipage}	
\centering
\includegraphics[width=0.22\textwidth]{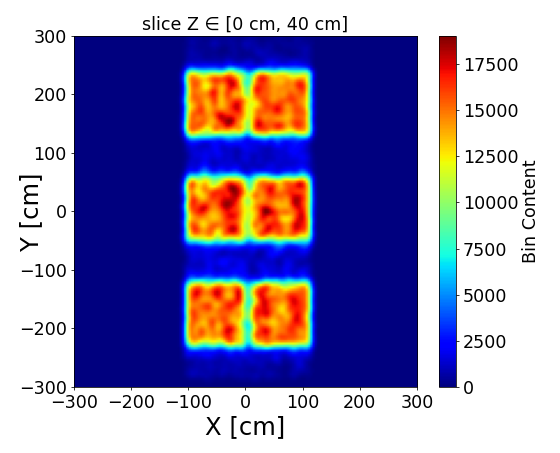} 
\includegraphics[width=0.22\textwidth]{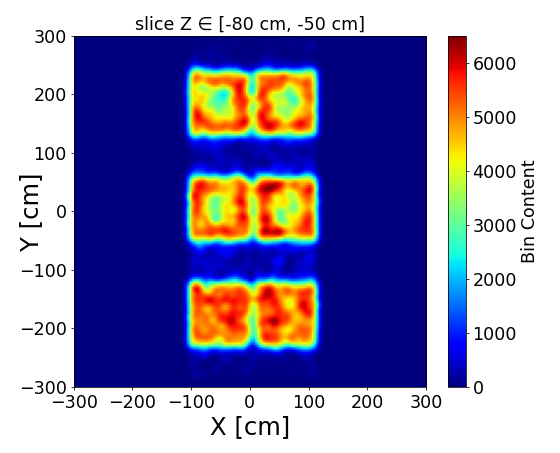}     
\caption{(a) \textit{XY} projection of the upper slice of the tomographic image, showing uniform distribution of cement bags across all pallets; (b) \textit{XY} projection of a lower slice, revealing distinct lower scattering density regions corresponding to concealed methamphetamine within the cement cargo.}
\label{fig:2d_cement_meth_reconstruction}
\end{figure}

To simulate the concealment of methamphetamine, a volume of the drug was embedded within cement bags in a GEANT4-based geometry. The simplified scenario models a standard 20-foot shipping container loaded with six pallets of cement. Of these, four pallets contain concealed cubic packages of methamphetamine, each measuring $50 \times 50 \times 50$ cm$^3$, hidden within the cement bags. The remaining two pallets consist solely of cement (see Figure~\ref{fig:cement_meth_geometry}(a)). Methamphetamine is an organic compound with the chemical formula $\mathrm{C_{10}H_{15}N}$. In most smuggling cases, it is encountered as methamphetamine hydrochloride, with the formula $\mathrm{C_{10}H_{15}N \cdot HCl}$ and an assumed density of 0.91\,g/cm$^3$~\cite{density}, reflecting the physical characteristics of methamphetamine hydrochloride powder commonly trafficked in illicit shipments. The surrounding material - cement was modeled with a bulk density of 1.4\,g/cm$^3$. Its chemical composition was approximated as follows: 61.66\% CaO, 19.83\% SiO$_2$, 2.32\% Fe$_2$O$_3$, 4.48\% Al$_2$O$_3$, 3.14\% MgO, and 2.57\% SO$_3$. 

The 5,000,000 muons were simulated, corresponding to a scanning duration of 5 minutes. Figure~\ref{fig:cement_meth_geometry}(b) shows a reconstructed image using the PoCA method with subsequent filtering for noise reduction. Segmenting the tomographic image into slices is necessary to enable visualization of hidden lower-density contraband within higher-density cargo. 
Figure\ref{fig:2d_cement_meth_reconstruction} displays the XY projections of two slices along the Z axis at different heights within the tomographic volume. In Figure~\ref{fig:2d_cement_meth_reconstruction}(a), the upper slice crosses cement bags, which appear as regions with uniform distributions across all six palettes. In the lower slice (~\ref{fig:2d_cement_meth_reconstruction}(b)), the concealed methamphetamine, despite being embedded within dense cement, appears as a distinct region of lower scattering density due to its lower material density, making it clearly distinguishable from the surrounding cement bags.

\subsection{Modeling high-Z special nuclear materials Hidden in Steel Pipes. } 

\begin{figure}[b]
\begin{minipage}{1.\linewidth}
\centering
\includegraphics[width=0.75\linewidth]{figures/ab.png}
\vspace{-0.5mm}  
\end{minipage}
\centering
\includegraphics[width=0.24\textwidth]{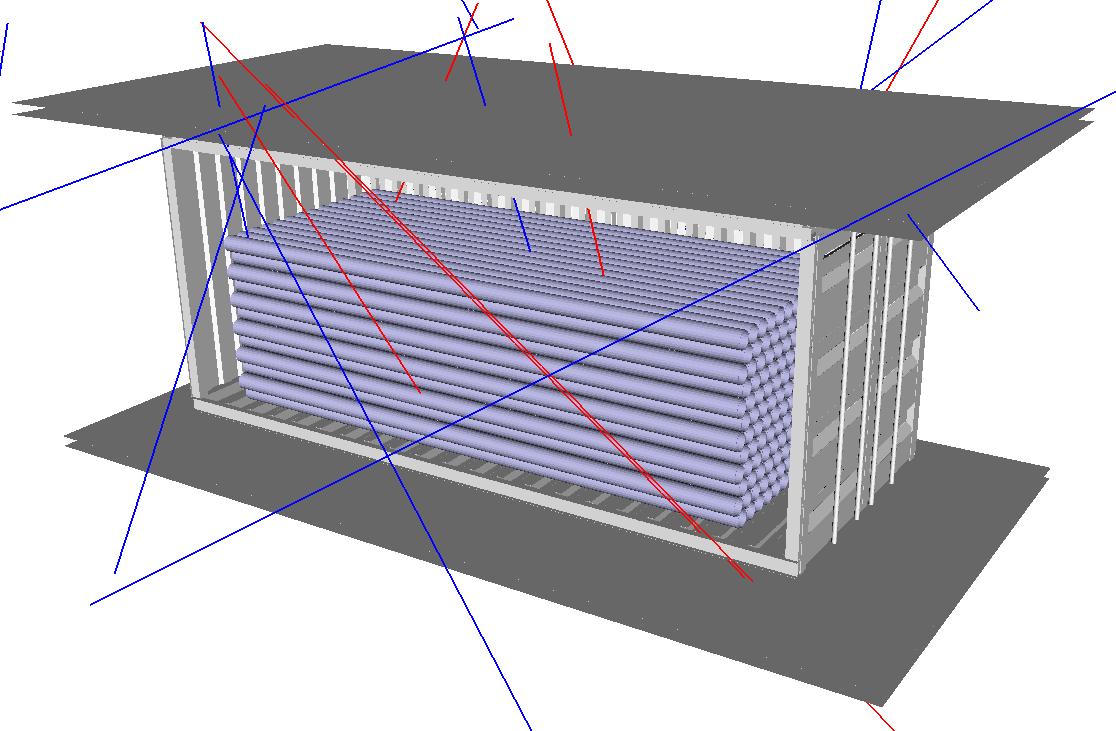}
\includegraphics[width=0.22\textwidth]{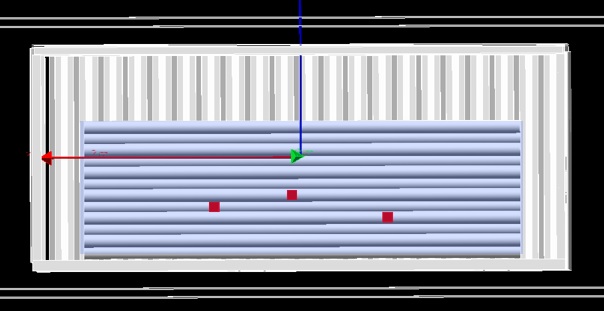}
\caption{A shipping container is loaded with steel pipes. Three 10 cm$^3$ cubes of SNM, represented in red, are concealed within the pipes.}
\label{fig:f20}
\end{figure}
\begin{figure*}[t]
\begin{minipage}{1.\linewidth}
\centering
\includegraphics[width=0.45\linewidth]{figures/ab.png}
\includegraphics[width=0.45\linewidth]{figures/cd.png}
\vspace{-0.5mm}  
\end{minipage}
\includegraphics[width=0.244\textwidth]{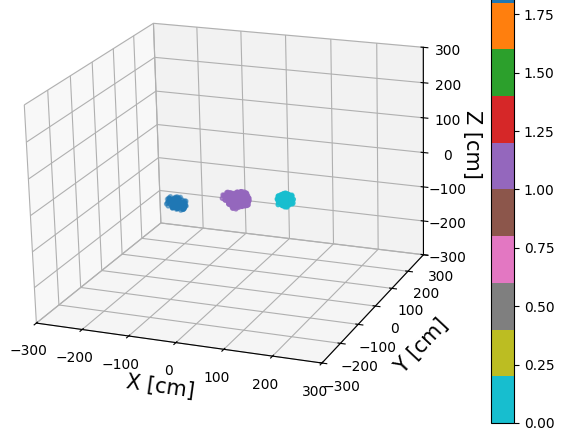}
\includegraphics[width=0.244\textwidth]{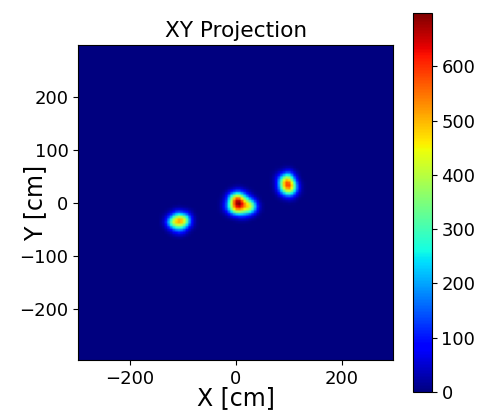}
\includegraphics[width=0.244\textwidth]{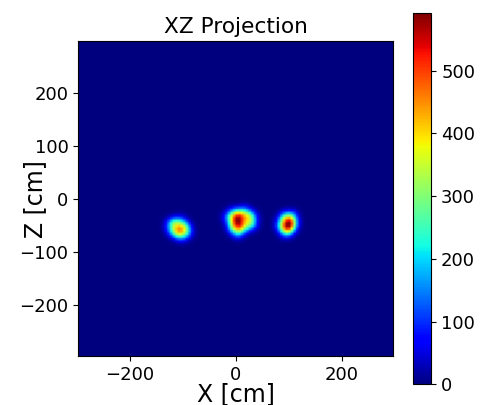}
\includegraphics[width=0.244\textwidth]{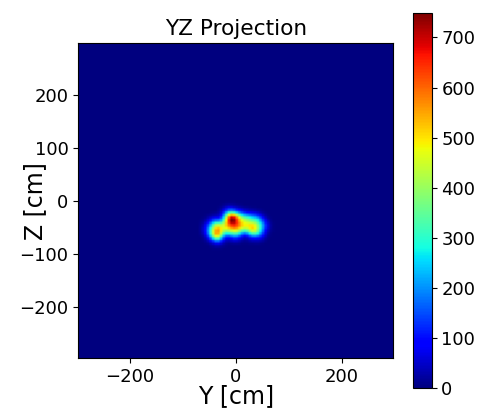}
\caption{(a) 3D visualization of three detected SNM cubes using the DBSCAN algorithm. The cubes are marked with different colors corresponding to the identified clusters.
\textit{XY}, \textit{YZ} and \textit{XZ} projections of tomographic image for 1 minute scanning time.}
\label{fig:f22}
\end{figure*}

Illicit trafficking of nuclear materials remains a critical security concern. We modeled a scenario where high-Z Special Nuclear Material (SNM) is concealed within steel pipes, common cargo items that can act as effective shielding. The GEANT4 model of container with pipes was created. We simulated one million muons sampled on the surface 10 $\times$ 10 cm$^2$ using CRY particle generator, which corresponding to a one-minute scan, to evaluate the system’s detection capability.
In this scenario, a shipping container is loaded with 166 steel pipes, each with a diameter of 16 cm, length of 500 cm and a thickness of 1 cm. Concealed within the pipes are three 10 $\times$ 10 $\times$ 10 cm$^3$ Special Nuclear Material (SNM) cubes (marked in red in Figure~\ref{fig:f20}), strategically hidden within the dense steel structure to evade detection.

The simulated data were processed using PoCA imaging algorithm. 
Performing image processing an adaptive thresholding algorithm was employed to suppress noise originating from the surrounding steel structures. This significantly improved image clarity and reduced false positives. Subsequently, the DBSCAN clustering algorithm was applied to automatically identify and localize the three SNM objects within the reconstructed volume.
Figure~\ref{fig:f22}(a) presents the reconstructed 3D image processed using the DBSCAN algorithm. Three SNM cubes are detected and marked with different colors corresponding to the identified clusters. Figures~\ref{fig:f22}(b)--(d) show the \textit{XY}, \textit{YZ}, and \textit{XZ} 2D projections, respectively. These views provide a clearer visualization of the SNM cubes hidden within the steel pipes. Noise from the surrounding steel structure has been effectively suppressed, enhancing both the image clarity and the detectability of SNM.

These results demonstrate the strong potential of real-time muon tomography combined with intelligent post-processing for the rapid detection of nuclear threats in dense and shielded cargo environments.

\section{Discussion and Conclusions}

This study demonstrates the effectiveness of a two-stage cargo screening system based on muon tomography for real-time cargo composition estimation. The integration of scattering and absorption measurements offers a highly sensitive, non-invasive method for analyzing a wide range of cargo types, with strong potential for enhancing customs screening procedures.

The proposed approach involves a rapid initial scan (10–20 seconds) to evaluate scattering and absorption rates, followed by a second stage in which a 3D image is reconstructed within 1 to 5 minutes. Post-processing techniques are then applied to detect and localize concealed contraband materials with high precision.

Monte Carlo-simulated scenarios confirm the system's ability to rapidly detect discrepancies between declared cargo content and measured physical properties. Categories of detected contraband include:
\begin{itemize}
    \item \textbf{Low-density, low-\textit{Z} materials}, commonly used to disguise the misdeclaration of high-value goods.
    \item \textbf{Hazardous and volatile substances}, which pose immediate safety risks and require prompt identification.
    \item \textbf{Special Nuclear Materials (SNM)}, often deliberately shielded within dense cargo structures to evade conventional detection—posing a critical non-proliferation concern.
\end{itemize}

Advanced 3D image processing techniques, including median and Gaussian filtering and adaptive thresholding, significantly improve image clarity by suppressing structural noise and enhancing anomaly detection. These methods have proven effective in revealing both low-\textit{Z} and high-\textit{Z} concealed threats, even when embedded in complex or dense cargo configurations.

The highly sensitive data processing algorithms developed in this work establish muon tomography as a powerful complement to traditional X-ray imaging systems. Especially in cases involving shielded or irregular cargo where X-rays may fail, muon-based systems offer a robust, high-accuracy, and non-destructive solution for modern border security and cargo inspection.

Future work should focus on integrating machine learning for anomaly detection and expanding real-world testing to further validate performance under operational conditions.

\begin{acknowledgments}
This work was partially supported by the EU Horizon 2020 Research and Innovation Programme under grant agreement no. 101021812 (“SilentBorder”).
\end{acknowledgments}

\section*{Data Availability Statement}
The data supporting this research are available from the corresponding author upon reasonable request.
\section{References}
\bibliography{aipsamp}

\end{document}